\documentclass[aps,prl,twocolumn,superscriptaddress,showpacs,floatfix,nobibnotes]{revtex4}
\usepackage{epsfig}
\usepackage{epstopdf}

\usepackage{graphicx}
\usepackage{longtable}
\usepackage{CJK}
\usepackage{color}

\usepackage{mathptmx, courier, pifont}
\usepackage[scaled=0.92]{helvet}
\usepackage[T1]{fontenc}
\usepackage{textcomp}

\begin{document}

\title{Understanding Xe isotopes near $A=130$ through the prolate-oblate shape phase transition}

\author{Wei Teng}
\affiliation{Department of Physics, Liaoning Normal University,
Dalian 116029, P. R. China}

\author{Sheng-Nan Wang}
\affiliation{Department of Physics, Liaoning Normal University,
Dalian 116029, P. R. China}

\author{Yu Zhang }\email{dlzhangyu_physics@163.com}
\affiliation{Department of Physics, Liaoning Normal University,
Dalian 116029, P. R. China}

\date{\today}

\begin{abstract}
A simple algebraic scheme incorporating the prolate-oblate shape phase transition (SPT) is proposed within the framework of the interacting boson model to describe the quadrupole deformation features of Xe isotopes near $A=130$. The analysis demonstrates that novel $\gamma$-soft modes, characterized by the unusual quadrupole moments $Q(2_1^+)<0$ and $0<Q(2_2^+)\ll |Q(2_1^+)|$, can emerge near the critical point of this SPT. This finding is further applied to interpret the properties of low-lying states in the relevant Xe nuclei, particularly the experimentally observed nearly vanishing spectroscopic quadrupole moment $Q(2_2^+)$, thereby offering new insights into the structure of a $\gamma$-soft deformed nucleus.
\end{abstract}

\maketitle

\begin{center}
\vskip.2cm\textbf{I. Introduction}
\end{center}\vskip.2cm

The diverse spectral structures observed in Xe isotopes offer a valuable opportunity for studying shape phase transitions (SPTs) and the evolution of collective modes in finite-$N$ systems~\cite{CJC2010,Casten2003}. In particular, Xe nuclei in the $A\approx130$ mass region have traditionally been regarded as approximate realizations~\cite{Casten1985,Brentano1988} of the O(6) dynamical symmetry (DS)~\cite{IachelloBook}. However, with the accumulation of experimental data~\cite{Kisyov2022,Morrison2020,Peters2019,Peters2016,Rainovski2010,Coquard2011,Coquard2010,Coquard2009}, especially concerning electromagnetic properties~\cite{Kisyov2022,Morrison2020}, their low-lying dynamics have been reinterpreted from various theoretical approaches~\cite{Kaneko2023,Nomura2021,Gupta2021,Budaca2020,Teruya2015,Hinohara2012,Meng2008,Bonatsos2006,Bonatsos2005} and multiple perspectives~\cite{Brentano1988,Peters2016,Coquard2009,Kaneko2023,Coquard2011,Clark2004}. For instance, since the E(5) critical point symmetry (CPS) of the U(5)-O(6) quantum phase transition (QPT) was proposed~\cite{Iachello2000}, $^{128}$Xe has been suggested as a candidate~\cite{Clark2004} for the E(5) CPS, following the initial empirical evidence from $^{134}$Ba~\cite{Casten2000}. Subsequent analyses~\cite{Coquard2009}, however, indicated that $^{128}$Xe may be not a close realization of the E(5) symmetry, leaving $^{130}$Xe as the most plausible E(5) candidate. A general conclusion is that in the Xe isotopes near $A=130$, symmetry higher than O(5) is more or less broken, whereas the O(5) symmetry remains relatively well preserved~\cite{Peters2016,Rainovski2010,Coquard2011}. Recently, Coulomb excitation measurements~\cite{Kisyov2022,Morrison2020} have provided more stringent tests of the nuclear shapes of Xe isotopes near $A=130$. These measurements reveal surprisingly large spectroscopic quadrupole moments for the $2_1^+$ and $4_1^+$ states, which appear to contradict the traditional O(6) prediction~\cite{IachelloBook} of vanishing quadrupole moments. More intriguingly, the data~\cite{Kisyov2022,Morrison2020} reveal small yet positive quadrupole moments $Q(2_2^+)>0$, which stand in contrast to the negative values $Q(2_1^+)<0$ and $Q(4_1^+)<0$ observed in these Xe isotopes. Notably, a nearly vanishing spectroscopic quadrupole moment of $Q(2_2^+)\simeq0.008$ eb has been reported for $^{128}$Xe~\cite{Kisyov2022}. This feature cannot be reproduced by either triaxial rotor models or microscopic shell-model calculations, posing a significant challenge to our current understanding of the quadrupole structure of the relevant Xe isotopes, as emphasized in \cite{Kisyov2022}.
Extending this observation to a broader context raises a question: what type of collective mode can give rise to such unusual spectroscopic quadrupole moments? Indeed, the importance of spectroscopic quadrupole moments in understanding O(6)-like or $\gamma$-unstable nuclei was previously highlighted in Ref.~\cite{Ostuka1993} through an analysis of $^{128}$Xe, where different theoretical calculations~\cite{Ostuka1993,Puddu1980} were shown to reproduce the level scheme similarly well while predicting significantly different quadrupole moments, even within the same theoretical framework.

On the theoretical side, the interacting boson model (IBM)~\cite{IachelloBook} offers a robust and versatile framework for describing a wide range of collective phenomena in atomic nuclei.
The O(6) DS of the IBM exhibits a $\gamma$-unstable ($\gamma$-soft) behavior at the mean-field level~\cite{IachelloBook}. This feature establishes a connection with the $\gamma$-soft model of Wilets and Jean (WJ)~\cite{WJ1956}, which was formulated within the framework of the geometric model. The O(6) type of $\gamma$-soft model, together with the $\gamma$-rigid rotor model of Davydov and Filippov (DF)~\cite{DF1958}, is frequently employed to explain the low-lying structures of nuclei exhibiting non-axially deformed shapes. It has been suggested that the two descriptions of triaxiality are, to some extent, equivalent for most observables~\cite{Otsuka1987,Zamfir1991}, despite differences in their geometric interpretations for $\gamma$ deformation. Furthermore, the O(6) limit is regarded as representing a critical point~\cite{Jolie2001} in the shape phase transition (SPT) between prolate and oblate nuclear shapes~\cite{Moreno1996}, which is described by the SU(3)-$\overline{\mathrm{SU(3)}}$ transition within the IBM~\cite{IachelloBook}. In this framework, both the SU(3) and $\overline{\mathrm{SU(3)}}$ limits are governed by the same two-body $QQ$ interactions, although different parametrization is used in constructing the quadrupole operator~\cite{Shirokov1998}. This results in to a sign change in the quadruple moments between the two SU(3) limits~\cite{Jolie2001}, indicating either a prolate or oblate-deformed ground state~\cite{Moreno1996}. The prolate-oblate paradigm has been shown to effectively explain the structural evolution observed in the Hf-Hg mass region~\cite{Jolie2003}, thereby offering strong empirical support for the shape phase diagram~\cite{Jolie2002} of the extended Casten triangle~\cite{Warner1983}. Recently, a shell model study~\cite{Kaneko2023} based on the quasi-SU(3) coupling mechanism~\cite{Zuker1995,Zuker2015} across the $N_\mathrm{n}=50$ and $82$ neutron shell gaps has been conducted to provide a microscopic understanding of why the SU(3)-O(6)-$\overline{\mathrm{SU(3)}}$ SPT occurs in nuclei~\cite{Jolie2001,Jolie2003}, with particular emphasis on the spectroscopic quadrupole moments of nuclei near $A=130$. On the other hand, an alternative algebraic description of the prolate-oblate SPT has been also proposed within the SU(3) limit~\cite{Zhang2012} of the IBM. This approach~\cite{Zhang2012} employs the three-body $QQQ$ interaction~\cite{Fortunato2011} or the third-order SU(3) Casimir operator, $\hat{C}_3\mathrm{[SU(3)]}$, instead of the conventional two-body $QQ$ interaction, to describe the oblate shape phase.
In this scheme, since the same quadrupole operator is used for both the prolate and oblate phases, the resulting prolate-oblate SPT pattern can even be analytically solved~\cite{Zhang2012}. Moreover, in this SU(3) formulation, the prolate phase and oblate phase correspond to the SU(3) irreducible representations (irreps)~\cite{Kotabook}, $(\lambda,\mu)=(2N,0)$ and $(\lambda,\mu)=(0,N)$, respectively, where $N$ denotes the total boson number, typically set to correspond to the number of valence nucleon pairs. This SU(3) scheme for the prolate-oblate SPT appears to be more closely aligned, in spirit, with the related SU(3) shell model approaches~\cite{Draayer1983,Draayer1989,Bonatsos2017I,Bonatsos2017II}, where the analogous relationship between $\gamma$ deformation and SU(3) irreps is preserved, except that the irreps in the shell model are determined mainly by the distribution of nucleons moving in specific shells.

In this work, we aim to explore the IBM framework~\cite{IachelloBook} to identify theoretical clues that may help understand the puzzling behavior of quadrupole moments in Xe isotopes near $A=130$. This is achieved by examining the critical features of the SU(3) IBM scheme~\cite{Zhang2012} in its description of the prolate-oblate SPT. The results are anticipated to offer a simplified perspective on the complex quadrupole structures exhibited by $\gamma$-soft nuclei.

\begin{center}
\vskip.2cm\textbf{II. The SU(3) Model for the Prolate-Oblate SPT }
\end{center}\vskip.2cm

The IBM Hamiltonian is constructed from two kinds of
boson operators: the $s$-boson with $l^\pi = 0^+$ and the
$d$-boson with $l^\pi=2^+$~\cite{IachelloBook}.
The total boson number $N$ is taken as the number of valence particle or hole pairs
for a given nucleus. All physical operators, including the Hamiltonian, are constructed utilizing the creation and annihilation operators of these bosons.

The IBM has three dynamical symmetry limits: U(5), O(6), and SU(3), which correspond to distinct nuclear shapes or collective modes.
Among these, the SU(3) limit is the primary focus of this work, as it is capable of describing both prolate and oblate rotational modes through the use of symmetry invariants~\cite{Zhang2012}, known as Casimir operators.
Specifically, the second- and third-order SU(3) Casimir operators are defined as~\cite{Kotabook}:
\begin{eqnarray}\label{C2}
&\hat{C}_2[\mathrm{SU(3)}]=2\hat{Q}\cdot\hat{Q}+\frac{3}{4}\hat{L}^2,\\
\label{C3}
&\hat{C}_3[\mathrm{SU(3)}]=-\frac{4\sqrt{35}}{9}(\hat{Q}\times\hat{Q}\times\hat{Q})^{(0)}
-\frac{\sqrt{15}}{2}(\hat{L}\times\hat{Q}\times\hat{L})^{(0)}\, ,
\end{eqnarray}
where $\hat{L}$ and $\hat{Q}$ denote the angular momentum and
quadrupole momentum operators defined in the SU(3) basis~\cite{IachelloBook},
\begin{eqnarray}\label{SQIL}
&&\hat{L}_u=\sqrt{10}(d^\dag\times\tilde{d})_u^{(1)}\, ,\\ \label{SQI}
&&\hat{Q}_u=(d^\dag\times\tilde{s}+s^\dag\times\tilde{d})_u^{(2)}-\frac{\sqrt{7}}{2}(d^\dag\times\tilde{d})_u^{(2)}\,
\end{eqnarray}
with the convention $\tilde{b}_u^l=(-1)^{l-u}b_{-u}^l$.
Eigenvalues of the Casimir operators can be expressed in terms of the SU(3) irreps with
\begin{eqnarray}\label{EC2}
&\langle\hat{C}_2[\mathrm{SU(3)}]\rangle=\lambda^2+\mu^2+3\lambda+3\mu+\lambda\mu,\\ \label{EC3}
&\langle\hat{C}_3[\mathrm{SU(3)}]\rangle=\frac{1}{9}(\lambda-\mu)(2\lambda+\mu+3)(\lambda+2\mu+3)\, ,
\end{eqnarray}
where the SU(3) irreps for a given $N$ are determined by
\begin{eqnarray}\label{irrep}\nonumber
(\lambda,\mu)&=&(2N,0),~(2N-4,2),~\cdots,~(0,N)~\mathrm{or}~(2,N-1),\\ \nonumber
&~&(2N-6,0),~(2N-10,2),~\cdots,\\
&~&\vdots\, .
\end{eqnarray}
Then, the allowed angular momentum quantum number can be extracted from the rule~\cite{IachelloBook,Elliott1958,Kotabook}:
\begin{eqnarray}\label{L}\nonumber
&&K=\mathrm{min}[\lambda,\mu],~\mathrm{min}[\lambda,\mu]-2,~\cdots,~0~\mathrm{or}~1\\
&&L=0,~2,~4,\cdots, ~\mathrm{max}[\lambda,\mu], ~\mathrm{for}~K=0,\\ \nonumber
&&L=K,~K+1,~\cdots, ~K+\mathrm{max}[\lambda,\mu],~\mathrm{for}~K>0,\\ \nonumber
&&M=-L,~-L+1,~\cdots,~L\, ,
\end{eqnarray}
in which $\mathrm{min}[a,b]$ ($\mathrm{max}[a,b]$) denotes the minimal (maximal) value between $a$ and $b$.

The $\gamma$ deformation in the SU(3) limit can be evaluated using the expression~\cite{Castanos1988}
\begin{eqnarray}\label{gamma}
\hat{\gamma}_{\mathrm{su3}}=\mathrm{tan}^{-1}\Big(\frac{\sqrt{3}(\hat{\mu}+1)}{2\hat{\lambda}+\hat{\mu}+3}\Big)\, .
\end{eqnarray}
In Eq.~(\ref{gamma}), $\hat{\lambda}$ and $\hat{\mu}$ should be treated as operators~\cite{Pan2003}, reducing to the usual $\lambda$ and $\mu$ values when
applied to the SU(3) limit. It follows that the prolate phase in the SU(3) limit is governed by $\langle-\hat{C}_2[\mathrm{SU(3)}]\rangle$, yielding
the ground-state SU(3) irrep $(\lambda,\mu)=(2N,0)$, whereas the oblate phase is dominated by $\langle\hat{C}_3[\mathrm{SU(3)}]\rangle$, leading to the ground-state irrep $(\lambda,\mu)=(0,N)$
for even $N$ or $(2,N-1)$ for odd $N$. For a boson system with $N=10$, this results in $\gamma_{\mathrm{su3}}\approx2^\circ$ for the prolate phase and $\gamma_{\mathrm{su3}}\approx56^\circ$ for the oblate phase. It should be noted that if the SU(3) quadrupole operator is redefined by replacing $-\sqrt{7}/2$ with $\sqrt{7}/2$~\cite{IachelloBook} in Eq.~(\ref{SQI}), the allowed SU(3) irrep $(\lambda,\mu)$ for a given $N$ will be obtained via $\lambda\leftrightarrow\mu$. In this way, the prolate and oblate phase can be accordingly described by $\langle-\hat{C}_3[\mathrm{SU(3)}]\rangle$ (or $\langle(\hat{Q}\times\hat{Q}\times\hat{Q})^{(0)}\rangle$) and $\langle-\hat{C}_2[\mathrm{SU(3)}]\rangle$ (or $\langle-\hat{Q}\cdot\hat{Q}\rangle$), respectively. Such a change between the prolate and oblate descriptions agrees with the mean-field analysis given in \cite{Fortunato2011}. In this study, the definition given in (\ref{SQI}) is consistently adopted.

To describe the prolate-oblate SPT, a schematic SU(3) Hamiltonian is formulated as~\cite{Zhang2012}
\begin{eqnarray}\label{HSU3}
&&\hat{H}_{\mathrm{SU(3)}}=-\frac{1}{N}\hat{C}_2[\mathrm{SU(3)}]+\frac{k}{N^2}\hat{C}_3[\mathrm{SU(3)}]\, ,
\end{eqnarray}
where $k$ serves as a control parameter. According to our previous analysis~\cite{Zhang2012}, the SU(3) system, as it evolves as a function of the control parameter, may undergo a first-order QPT between the prolate and oblate phases, arising from level crossing. It is evident that the Hamiltonian under consideration accounts exclusively for intrinsic excitations~\cite{Kirson1985}. In the context of an SU(3) system, rotational bands are classified by the quantum numbers $K$ and $(\lambda,\mu)$. Here, intrinsic excitation means that only the band head energy of a collective rotational band can be explicitly determined by the Hamiltonian (\ref{HSU3}), whereas all band members corresponding to a given SU(3) irrep $(\lambda,\mu)$ are energetically degenerate. Within the mean-field approximation, such intrinsic excitations can be associated with $\beta$- or $\gamma$-vibrational modes~\cite{Leviatan1985}. Given that the Hamiltonian (\ref{HSU3}) does not depend on angular momentum $L$, the excitation energy function $E(k,\lambda,\mu)$ can be analytically derived based on Eq.~(\ref{EC2})-(\ref{irrep}). Rotational bands can be generated by incorporating the rotational term $\hat{L}^2$ into the Hamiltonian when necessary. This may lead to an energy sequence proportional to $L(L+1)$ for a given $K$, as detailed in Eq.~(\ref{L}). To maintain clarity, we continue to employ the Hamiltonian form given in Eq.~(\ref{HSU3}) when discussing the phase transition between prolate and oblate intrinsic configurations. Accordingly, the critical point (level crossing point) $k_\mathrm{c}$ is determined by the equations:
$E_g(k_\mathrm{c},2N,0)=E_g(k_\mathrm{c},0,N)$ for even $N$ and $E_g(k_\mathrm{c},2N,0)=E_g(k_\mathrm{c},2,N-1)$ for odd $N$.
It can be analytically derived that the value of $k_\mathrm{c}$ is given by
\begin{eqnarray}\label{kc}
k_\mathrm{c}=\frac{3N}{2N+3}\,
\end{eqnarray}
for both even $N$ and odd $N$. In fact, at this critical point, not only do the ground-state energies satisfy the above equations, but all energy levels $E_g(k_\mathrm{c},2N-2t,t)$ with $t=0,~2,~4,~\cdots,N$ or $N-1$ become degenerate, indicating a complete level crossing~\cite{Zhang2012}. Consequently, the states belonging to the SU(3) irreps satisfying $\lambda+2\mu=2N$ are lowered to the lowest energy levels, becoming energetically separated from the other SU(3) irreps~\cite{Zhang2012}.

To calculate $B(E2)$ transitions and spectroscopic quadrupole moments, the $E2$ operator is chosen as
\begin{eqnarray}\label{E2}
\hat{T}(E2)=e\hat{Q}\, ,
\end{eqnarray}
where $e$ denotes the effective charge and the $Q$ operator is taken to be the same as that used in the Hamiltonian (see Eq.~(\ref{SQI})).
The $B(E2)$ transitions and quadrupole moments can then be evaluated using the following expressions:
\begin{eqnarray}
B(E2;L_i\rightarrow L_f)=\frac{|\langle\alpha_fL_f\parallel \hat{T}(E2)\parallel\alpha_iL_i\rangle|^2}{2L_i+1}\,
\end{eqnarray}
and
\begin{equation}
Q(L)=\sqrt{\frac{16\pi}{5}}\langle\alpha LM|\hat{T}(E2)|\alpha LM\rangle|_{M=L}\, ,
\end{equation}
where $\alpha$ represents all relevant the quantum numbers other than $L$ and $M$.

\begin{figure}
\begin{center}
\includegraphics[scale=0.3]{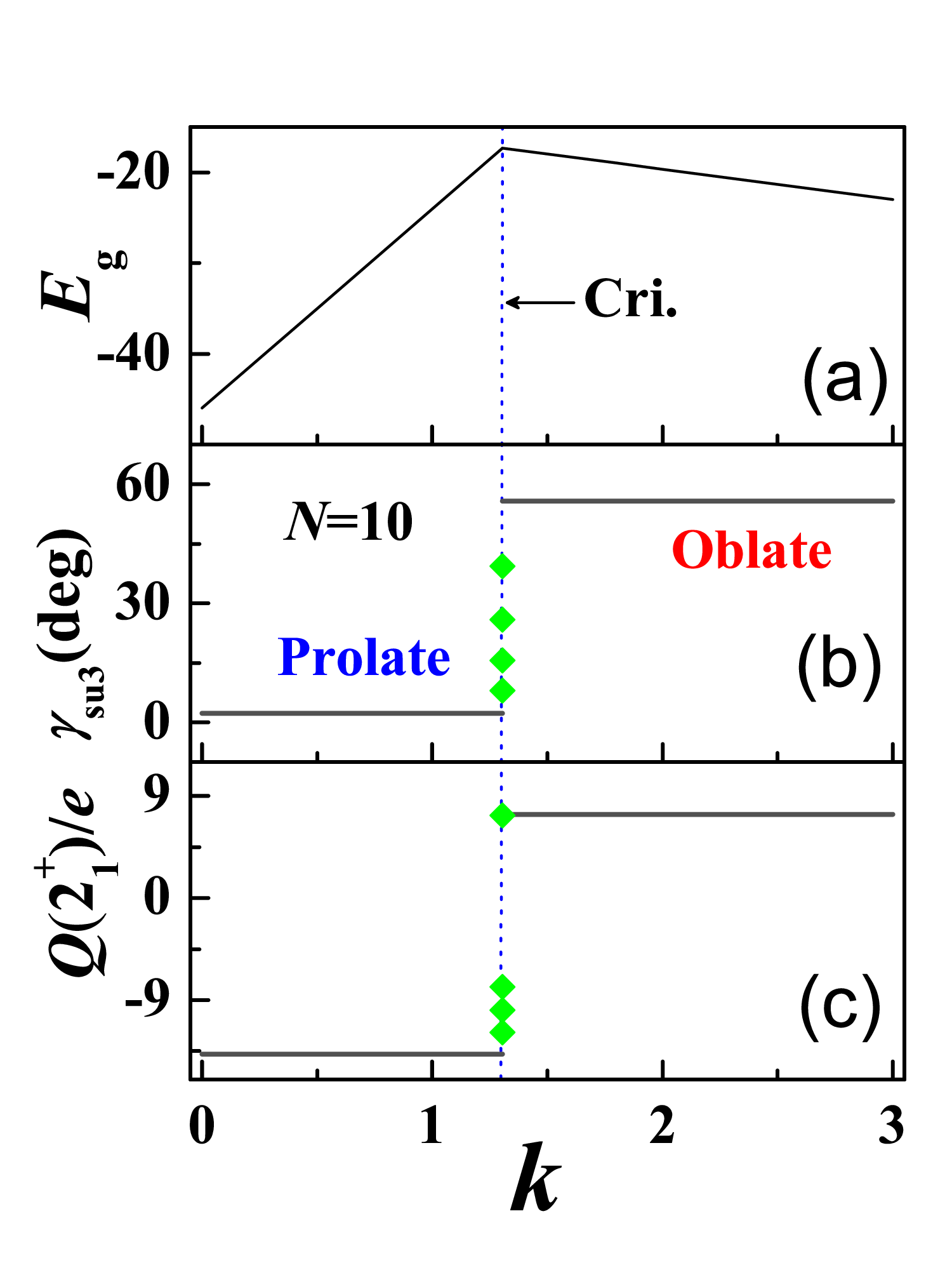}
\caption{(a) The ground-state energy $E_\mathrm{g}$ evolves as a function of $k$, with the critical point $k_\mathrm{c}=3N/(2N+3)$ indicated by "Cri". (b) The same as in (a), but for the $\gamma$ deformations evaluated using Eq.~(\ref{gamma}), where the green scatter points at the critical point represent the results for the degenerate SU(3) irreps $(\lambda,\mu)=(N-2t,t)$ with $t=2,~4,~6,~8$. (c) The same as in (b), but for the spectroscopic quadrupole moment $Q(2_1^+)$, where the $2_1^+$ state is assumed to originate from the $K=0$ band within a given $(\lambda,\mu)$. All results are obtained by solving the SU(3) Hamiltonian (\ref{HSU3}) for $N=10$. \label{F1}}
\end{center}
\end{figure}

\begin{figure}
\begin{center}
\includegraphics[scale=0.2]{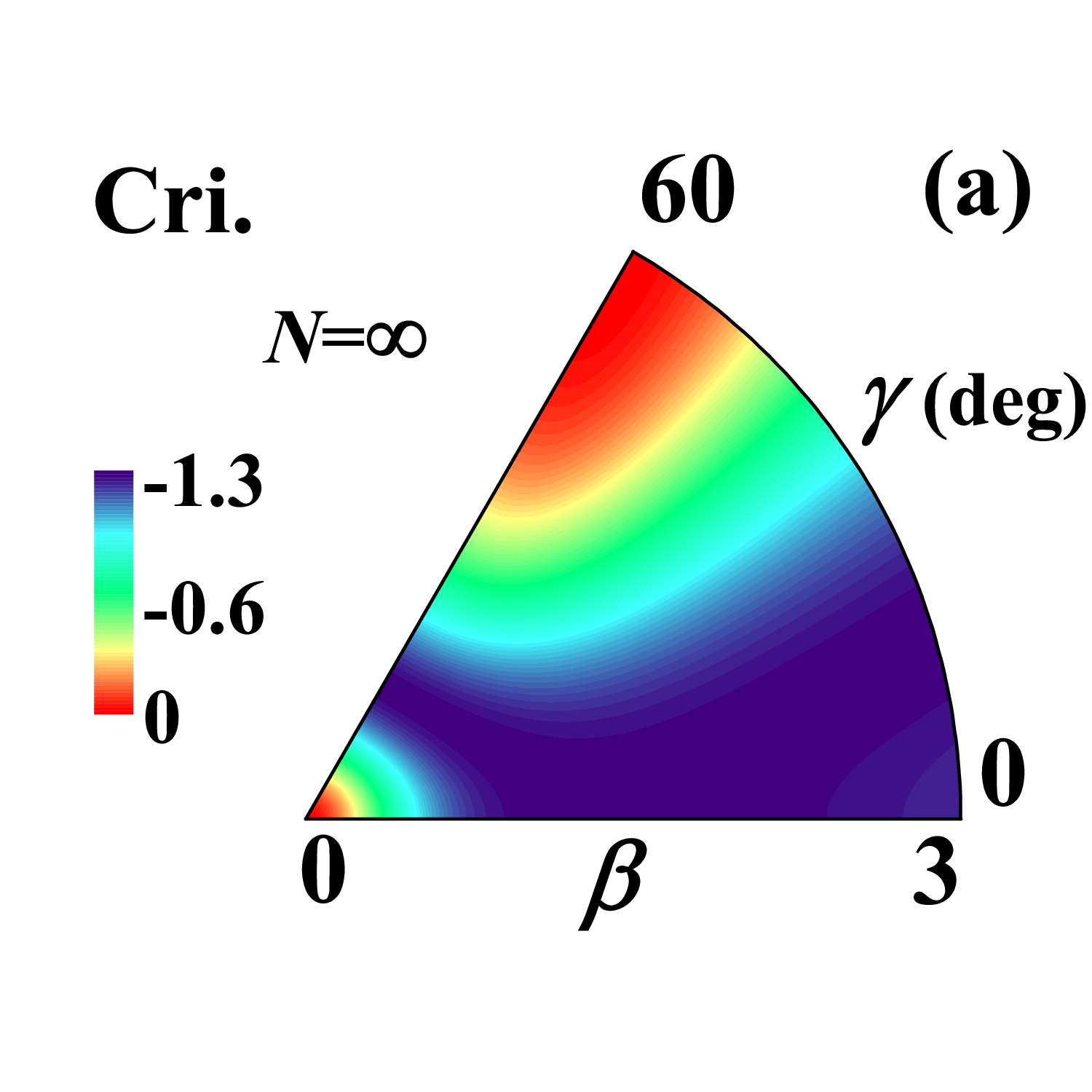}
\includegraphics[scale=0.2]{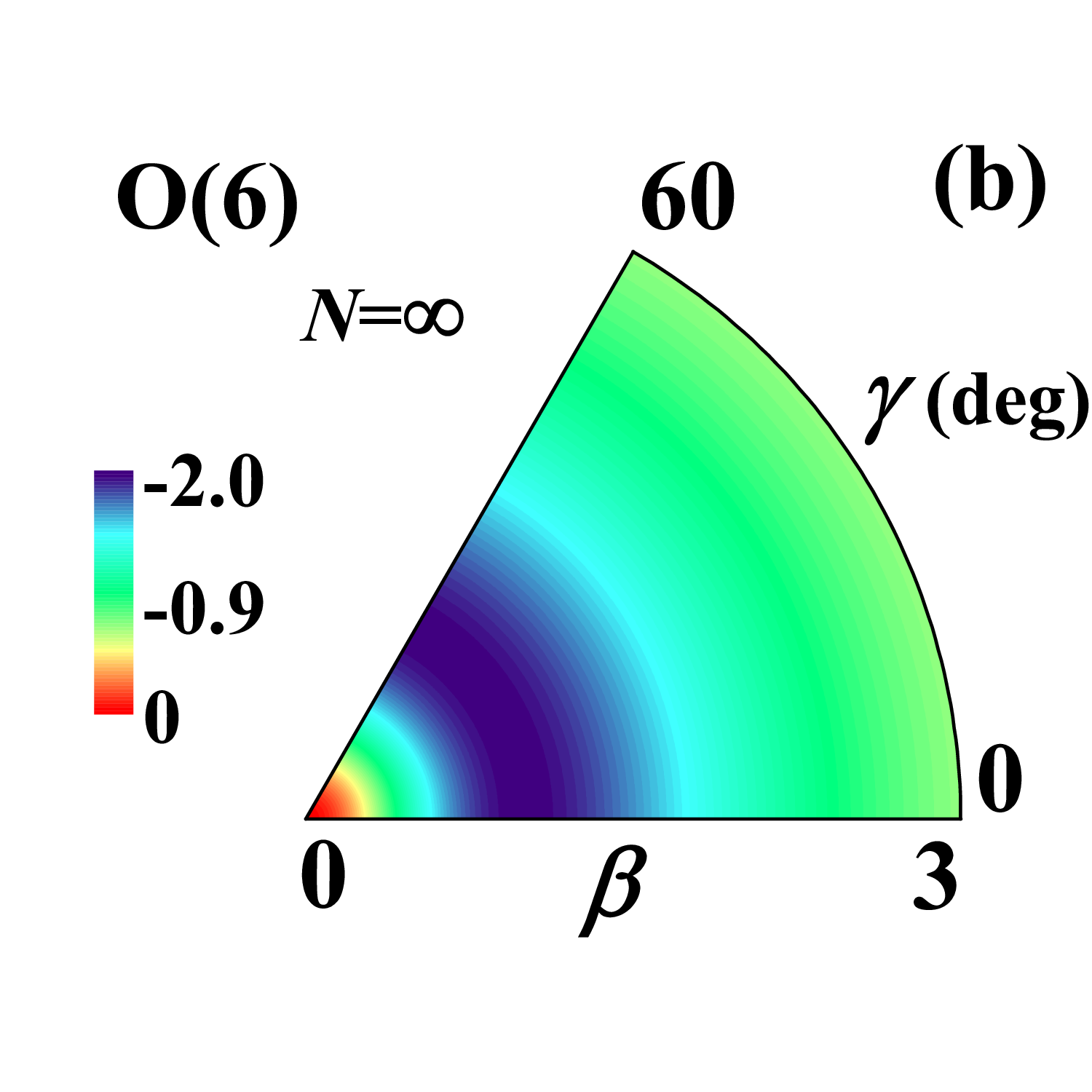}
\caption{The contour plots for the classical potentials $V(\beta,\gamma)$, obtained at the critical point of the prolate-oblate SPT and in the O(6) limit (see the text for parameters), using the coherent-state method. \label{F2}}
\end{center}
\end{figure}

\begin{figure}
\begin{center}
\includegraphics[scale=0.3]{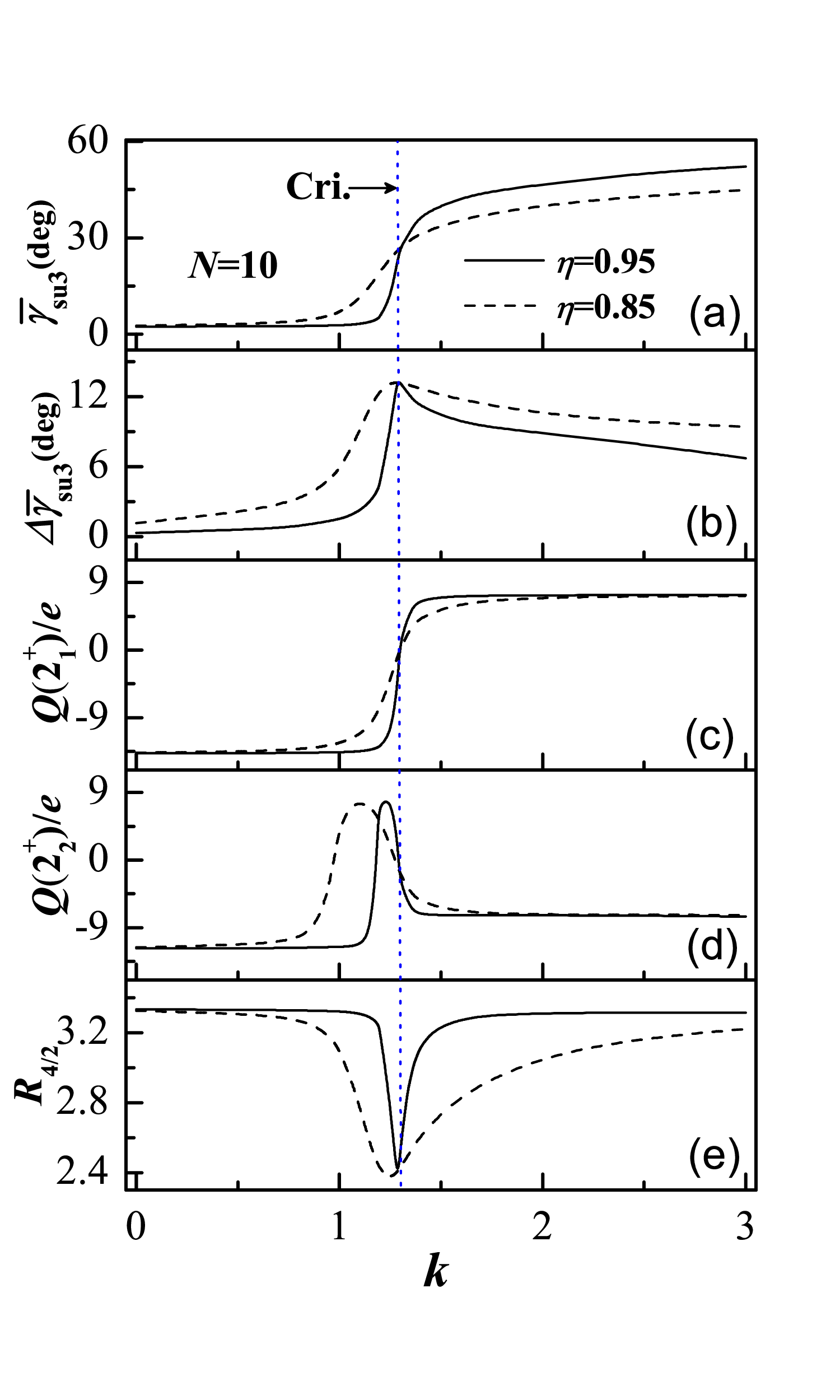}
\caption{(a) The evolutions of the $\gamma$ deformation, solved from the Hamiltonian (\ref{H}) as a function of $k$, is shown for $\eta=0.95$ and $\eta=0.85$, respectively. (b) The same as in (a), but for
the $\gamma$ fluctuations. (c) The same as in (a), but for $Q(2_1^+)/e$ . (d) The same as in (a), but for $Q(2_2^+)/e$. (e) The same as in (a), but for the energy ratio $R_{4/2}$. The total boson number used in the calculations is fixed at $N=10$, and "Cri." denotes the critical point $k_\mathrm{c}=3N/(2N+3)$.  \label{F3}}
\end{center}
\end{figure}

\begin{figure}
\begin{center}
\includegraphics[scale=0.3]{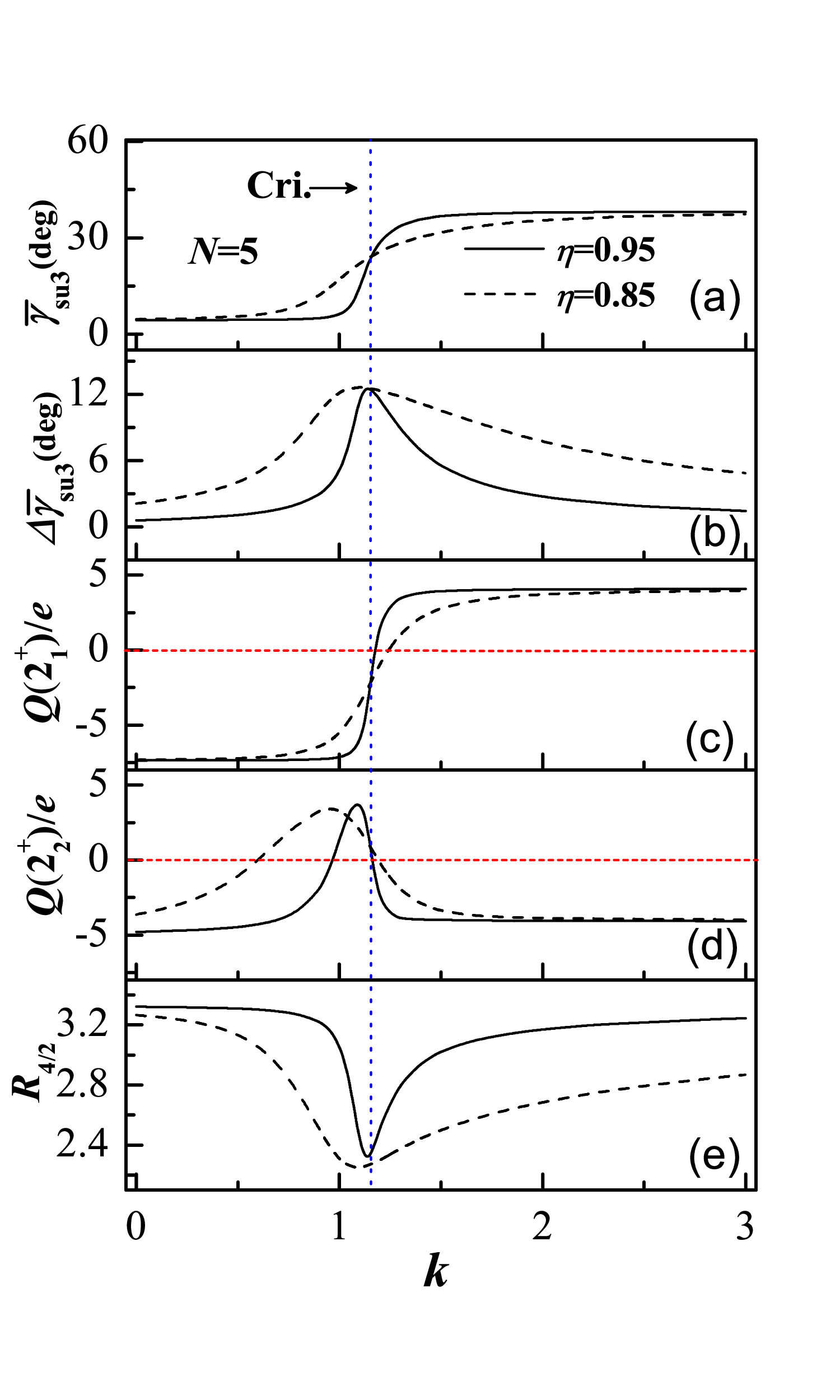}
\caption{The same as in Fig.~\ref{F3}, but for $N=5$, with parallel dashed lines added to
indicate where $Q(2_{1,2}^+)=0$. \label{F4}}
\end{center}
\end{figure}

To offer a clear illustration of the prolate-oblate SPT, we present in Fig.~\ref{F1} the evolutions of the ground-state energy $E_\mathrm{g}$, the spectroscopic quadrupole moment $Q(2_1^+)$, and the ground-state $\gamma$ deformation evaluated using Eq.~(\ref{gamma}). The results ($N=10$) indicate that the system, as it evolves as a function of $k$, indeed undergoes a prolate to oblate SPT, with the ground-state $\gamma$ deformation (order parameter) exhibiting a jump at the critical point $k_\mathrm{c}$ from $\gamma_{\mathrm{su3}}\approx0^\circ$ to $\gamma_{\mathrm{su3}}\approx60^\circ$, which signifies a first-order phase transition. The nature of the phase transition is further supported by the observation that the ground-state energy, as a function of the control parameter $k$, remains continuous at $k_\mathrm{c}$, while its first derivative shows a discontinuity, as indicated by the results presented in Fig.~\ref{F1}(a). As illustrated in Fig.~\ref{F1}(c), this SPT is corroborated by the evolution of the spectroscopic quadrupole moment $Q(2_1^+)$, which abruptly changes from a negative constant to a positive value as the system crosses the critical point $k=k_\mathrm{c}$. Collectively, these findings confirm that prolate-oblate SPT in this scheme is clearly well defined at finite $N$~\cite{Zhang2012}.

As previously discussed, all states associated with the SU(3) irreps $(\lambda,\mu)=(2N-2t,t)$, where $t=0,~2,~4,~\cdots$, become energetically degenerate at the critical point $k_\mathrm{c}$. As further seen in Fig.~\ref{F1}(b) and Fig.~\ref{F1}(c), although these irreps are degenerate, they correspond to different values of $\gamma_{\mathrm{SU(3)}}$ and exhibit varying spectroscopic quadrupole moments, which indicates the presence of intrinsic instabilities in the quadrupole deformation of the critical system. To understand the mean-field geometry of the critical point system, one can examine its large-$N$ (classical) limit by employing the coherent state defined as~\cite{IachelloBook}
\begin{eqnarray}\label{coherent}
|\beta, \gamma, N\rangle=A[s^\dag + \beta \mathrm{cos} \gamma~
d_0^\dag\ + \frac{1}{\sqrt{2}} \beta \mathrm{sin} \gamma (d_2^\dag +
d_{ - 2}^\dag)]^N |0\rangle\,
\end{eqnarray}
with $A=1/\sqrt{N!(1+\beta^2)^N}$. The classical potential per boson for a given Hamiltonian in the large-$N$ limit can be expressed as a function of the deformation parameters $\beta$ and $\gamma$ by
\begin{eqnarray}\label{V}
V(\beta,\gamma)=\frac{1}{N}\langle \beta, \gamma, N|\hat{H}|\beta, \gamma, N\rangle_{N\rightarrow\infty}\, .
\end{eqnarray}
Specifically, the classical potential corresponding to the Hamiltonian (\ref{HSU3}) at the critical point
is presented in Fig.~\ref{F2}, where the potential for the O(6) limit is defined as $V_{\mathrm{O(6)}}=\frac{1}{N}\langle \beta, \gamma, N|-2\hat{Q}^\prime\cdot\hat{Q}^\prime/N|\beta, \gamma, N\rangle_{N\rightarrow\infty}$ with $\hat{Q}^\prime=(d^\dag s+s^\dag\tilde{d})^{(2)}$, and is also provided for comparison. Notably, the critical point of the prolate-oblate SPT at the mean-field level is identified as $k_\mathrm{c}=1.5$~\cite{Fortunato2011}, which agrees well with the analytical expression given in (\ref{kc}) obtained from the level crossing. As observed from Fig.~\ref{F2}(a), the potential at the critical point exhibits a $\gamma$-soft behavior, with the equilibrium values of $\gamma$ ranging within $\gamma_\mathrm{e}\in[0^\circ,60^\circ]$, resembling the O(6) potential shown in Fig.~\ref{F2}(b). However, in contrast to the fixed equilibrium value of $\beta$ in the O(6) limit, the $\gamma$-soft potential at the critical point also displays evident $\beta$-softness on the prolate side~\cite{Fortunato2011}.

Although the mean-field analysis can only provide a qualitative reference for realistic cases at finite-$N$, the results clearly demonstrate that the current prolate-oblate SPT scheme can yield a $\gamma$-soft structure at the critical point, similar to the original SU(3)-O(6)-$\overline{\mathrm{SU(3)}}$ scheme~\cite{Jolie2001}. Apart from that, the $\gamma$ softness exhibited by the critical point system in the large-$N$ limit is consistent with the finite-$N$ results, which show multiple $\gamma$ deformations coexisting (i.e., being degenerate in energy) at the critical point, as indicated by the results shown in Fig.~\ref{F1}(b). It is thus concluded that $\gamma$-soft dynamics may naturally emerge from the competition (phase transition) between two types of $\gamma$-rigid SU(3) rotor modes, in contrast to the traditional approach that incorporates an O(6)-like term at the Hamiltonian level~\cite{Zhang2013,Bonatsos2024}. This may suggest an alternative perspective on the origin of $\gamma$-soft modes in an identical-particle system, particularly given that the SU(3) symmetry can be similarly realized in an identical-fermion system~\cite{Kotabook,Bonatsos2017II}. It would be of interest to further examine this point within other models, which, however, lies beyond the scope of the present study.

For more general situations, the model Hamiltonian incorporating the prolate-oblate SPT is designed as
\begin{eqnarray}\label{H}
\hat{H}=\varepsilon\Big[\frac{\eta}{8}\hat{H}_{\mathrm{SU(3)}}+(1-\eta)\hat{n}_d\Big]\,
\end{eqnarray}
with $\hat{n}_d=\sum_m d_m^\dag d_m$.
In Eq.~(\ref{H}), $\eta$ serves an additional control parameter alongside $k$ from Eq.~(\ref{HSU3}), and $\varepsilon$ acts as a scaling factor, which
will be fixed at $\varepsilon=1.0$ unless otherwise stated. The inclusion of the $\hat{n}_d$ term fulfills two purposes:
it removes degeneracies arising from the SU(3) Hamiltonian (\ref{HSU3}) and introduces contributions from the vibrator mode (the U(5) limit), making the model applicable to realistic scenarios.
Due to the inclusion of the U(5) mode, the Hamiltonian (\ref{H}) can describe a shape phase diagram analogous to the Casten triangle~\cite{Jolie2002}, which is commonly modeled using the well-known consistent-$Q$ Hamiltonian~\cite{Warner1983} in the IBM. A detailed mean-field analysis of this shape phase diagram and its extension was presented in \cite{Fortunato2011}, revealing a small region exhibiting triaxiality within the model. Additionally, a numerical investigation into the evolution of different observables across the prolate-oblate SPT was reported in \cite{Wang2023}, following the mean-field analysis given in \cite{Fortunato2011} and analytical analysis provided in \cite{Zhang2012}. In what follows, we focus on specific features of the prolate-oblate SPT described by the Hamiltonian (\ref{H}), leaving a systematic analysis of other SPTs for future work.

Note that QPTs are rigorously defined only in the large-$N$ limit. According to the mean calculations of a similar Hamiltonian form, as reported in \cite{Fortunato2011}, the inclusion of the $\hat{n}_d$ term in (\ref{H}) may result in the formation to a narrow region that allows for triaxiality when $\eta<1$, which is further confirmed by the numerical analysis using the coherent-state method. Within this region, the prolate-oblate transitions along the $k$-axis may split into two consecutive second-order phase transitions: the prolate-triaxial (A) and triaxial-oblate (B) SPTs. The corresponding critical points, $k_\mathrm{c}$, become functions of $\eta$. Although deriving an analytical expression for $k_\mathrm{c}(\eta)$ is challenging, mean-field analysis indicates that the two critical points $k_\mathrm{c}^{\mathrm{A}}$ and $k_\mathrm{c}^{\mathrm{B}}$ are generally very close to each other and tend to converge toward the value in the SU(3) limit obtained at $\eta=1$. For instance, mean-field calculations (in the large-$N$ limit) for the current Hamiltonian yield $k_\mathrm{c}^{\mathrm{A}}\approx1.33$ and $k_\mathrm{c}^{\mathrm{B}}\approx1.36$ for $\eta=0.8$, which are close to the critical value $k_\mathrm{c}=1.5$ observed at $\eta=1$. Therefore, for finite $N$, where the QPT is not rigorously defined except at $\eta=1$, it is reasonable to approximate $k_\mathrm{c}=3N/(2N+3)$ as the critical point of the prolate-oblate transition in systems near SU(3) limit. In addition, by including the $\hat{n}_d$ term, the Hamiltonian (\ref{H}) may describe a first-order phase transition along the $\eta$-axis at the mean-field level, from a spherical to a deformed shape (either prolate or oblate, depending on the value of $k$). Using the coherent-state method, the critical point $\eta_\mathrm{c}(k)$ can be determined numerically, yielding $\eta_\mathrm{c}\in[\frac{8}{17},\frac{1}{2})$ for first-order spherical-prolate SPTs when $k\in[0,\frac{9}{8})$, and $\eta_\mathrm{c}\in(0,\frac{1}{2})$ for first-order spherical-oblate SPTs when $k>\frac{9}{8}$. The point $(\eta,k)=(\frac{1}{2},\frac{9}{8})$ corresponds to a triple point, where the spherical-deformed SPT becomes second-order. Notably, as $k$ increases toward infinity, the critical points of the spherical-oblate SPTs approach the U(5) limit ($\eta=0$). Based on this analysis, it follows that a prolate-oblate transition can occur only if $k>\frac{1}{2}$ for the system described by the current Hamiltonian.

To illustrate the $\gamma$ deformation in a finite $N$ system, using formula (\ref{gamma}) is more convenient than employing the coherent state method. It provides average values of the $\gamma$ deformation, $\overline{\gamma}_{\mathrm{su3}}=\langle\hat{\gamma}_{\mathrm{su3}}\rangle$,
for cases that deviate from the SU(3) symmetry, thus representing dynamical $\gamma$ deformation. Based on this formulation, $\gamma$-softness can be examined using the fluctuation defined by
\begin{eqnarray}\label{deltagamma}
\Delta\overline{\gamma}_{\mathrm{su3}}=\sqrt{\langle\hat{\gamma}_{\mathrm{su3}}^2\rangle-\langle\hat{\gamma}_{\mathrm{su3}}\rangle^2}\, ,
\end{eqnarray}
where $\langle \hat{A}\rangle$ denotes the expectation value of $\hat{A}$ in a given state $|\Psi\rangle_{\xi L M}$. In the calculation of $\gamma$ fluctuations, the state vector should be expanded in terms of the orthogonal SU(3) basis in the IBM~\cite{IachelloBook},
\begin{eqnarray}
|\Psi\rangle_{\xi L M}=\sum_{\lambda\mu\chi}C_{\lambda\mu\chi}^\xi|N,(\lambda,\mu),\chi,LM\rangle_{\mathrm{su3}}\, ,
\end{eqnarray}
where $C_{\lambda\mu\chi}^\xi$ are the expansion coefficients, with $\xi$ representing all quantum numbers other than $L$ and $M$, and $\chi$ corresponds to the additional quantum number in the reduction of SU(3) irreps. By applying Eq.~(\ref{gamma}), it follows that
\begin{eqnarray}
&&\langle\hat{\gamma}_{\mathrm{su3}}\rangle_{\xi}=\sum_{\lambda\mu\chi}|C_{\lambda\mu\chi}^\xi|^2\langle~\hat{\gamma}_{\mathrm{su3}}~\rangle_{\mathrm{su3}},\\ &&\langle\hat{\gamma}_{\mathrm{su3}}^2\rangle_{\xi}=\sum_{\lambda\mu\chi}|C_{\lambda\mu\chi}^\xi|^2\langle~\hat{\gamma}_{\mathrm{su3}}^2~\rangle_{\mathrm{su3}}\, ,
\end{eqnarray}
where $\langle~\hat{\gamma}_{\mathrm{su3}}^n~\rangle_{\mathrm{su3}}=\Big[\mathrm{tan}^{-1}\Big(\frac{\sqrt{3}(\mu+1)}{2\lambda+\mu+3}\Big)\Big]^n$ with $n=1,~2$.
With these expressions, one can evaluate both $\bar{\gamma}_{\mathrm{su3}}$ and $\Delta\overline{\gamma}_{\mathrm{su3}}$ for any given state.

To identify the critical behaviors in the prolate-oblate SPT, we examine the evolutions of the average $\gamma$ deformation ($\bar{\gamma}_{\mathrm{su3}},~\Delta\overline{\gamma}_{\mathrm{su3}}$) for the ground states, spectroscopic quadrupole moments for two $2^+$ states, and the energy ratio $R_{4/2}=E(4_1^+)/E(2_1^+)$. These quantities are key to characterizing the low-lying structures of Xe nuclei near $A=130$~\cite{Kisyov2022,Morrison2020}. Results for $N=10$ and $N=5$, obtained from two cases ($\eta=0.95$ and $\eta=0.85$), are presented in Fig.~\ref{F3} and Fig.~\ref{F4}, respectively, to evaluate the effects of finite-$N$ correction and vibrational mode on the SPT. As observed from Fig.~\ref{F3}(a), the $\gamma$ deformation as a function of $k$ may monotonically increase from $\bar{\gamma}_{\mathrm{su3}}\approx0^\circ$ to $\bar{\gamma}_{\mathrm{su3}}>40^\circ$. Although a discontinuous change in the $\gamma$ deformation, such as that observed in Fig.~\ref{F1}(b), does not occur here, the rapid change in $\bar{\gamma}_{\mathrm{su3}}$ around $k_\mathrm{c}=3N/(2N+3)$ suggests that the fundamental characteristics of prolate-oblate SPT are well preserved even in systems deviating from SU(3) limit. Therefore, it is reasonable to regard $k_\mathrm{c}=3N/(2N+3)$ as the critical point of the transition, even though QPT is not strictly defined at finite $N$, when moving away from SU(3) limit.

Apart from the finite-$N$ effect, the results indicate that an increasing contribution of $\hat{n}_d$ may, to some extent, smooth out the SPT features. Furthermore, as shown in Fig.~\ref{F3}(b), fluctuations in $\bar{\gamma}$ reach a maximum and peak around $k_\mathrm{c}$, indicating that $\gamma$ softness is maximized in critical point systems. This result is consistent with the mean-field picture of the critical point illustrated in Fig.~\ref{F2}(a). As observed in Fig.~\ref{F3}(c), the results for the spectroscopic quadrupole moment of the $2_1^+$ state support the occurrence of prolate-oblate SPT around $k_\mathrm{c}$, where the values of $Q(2_1^+)$ rapidly transition from negative to positive. In contrast, the values of $Q(2_2^+)$ remain negative in both prolate and oblate phases but exhibit a roller coaster-like variation across $Q(2_2^+)=0$ near the critical point. As further seen in Fig.~\ref{F3}(e), the evolution of the $R_{4/2}$ ratio may also display a clear critical behavior in this prolate-oblate SPT. Compared to large values observed in either the prolate or oblate phases, where $R_{4/2}\simeq3.3$, as expected for an axial rotor, the ratio decreases significantly near the critical point down to $R_{4/2}<2.4$, which provides an experimental criterion for identifying the critical point structures. The substantial reduction in the $R_{4/2}$ ratio, together with the observed large $\gamma$-softness, indicates that quadrupole structure emerging at critical points represents a novel mode that is fundamentally distinct from a rigid rotor or other intermediate cases between two axial rotors (prolate and oblate) traditionally defined within the SU(3) symmetry~\cite{Kotabook}.

The similar critical behaviors of different quantities presented in Fig.~\ref{F3} can be identified even at smaller $N$, as demonstrated in Fig.~\ref{F4}, where results for $N=5$ are shown to offer more relevant references for realistic situations, since the boson number is closer to those of the Xe nuclei near $A=130$. A particularly interesting observation in this case is that the critical point system at $k_\mathrm{c}$ may yield $Q(2_1^+)<0$ and $Q(2_2^+)\sim0$, as indicated, for example, by the results for $\eta=0.85$ shown in Fig.~\ref{F4}(c) and Fig.~\ref{F4}(d). This feature aligns well with the exotic quadrupole moments observed in the related Xe isotopes~\cite{Kisyov2022,Morrison2020}. It should be emphasized again that the critical point $k_\mathrm{c}=3N/(2N+3)$ in the present scheme is identified from level crossings in the SU(3) limit, rather than $Q(2_1^+)=0$ or $\bar{\gamma}_{\mathrm{su3}}=30^\circ$.

\begin{center}
\vskip.2cm\textbf{III. Comparison with $^{128,130,132}$Xe and Discussions}
\end{center}\vskip.2cm

\begin{figure}
\begin{center}
\includegraphics[scale=0.35]{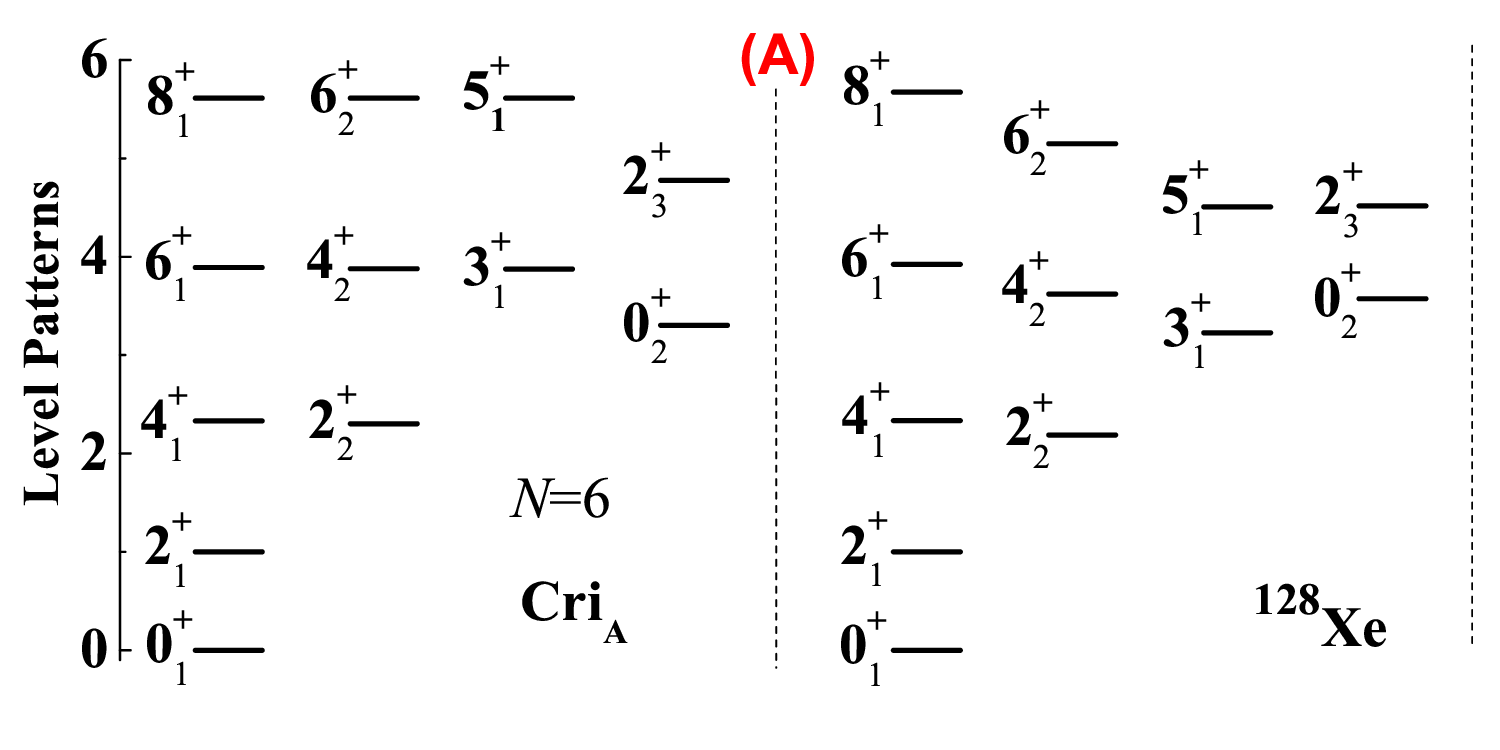}
\includegraphics[scale=0.35]{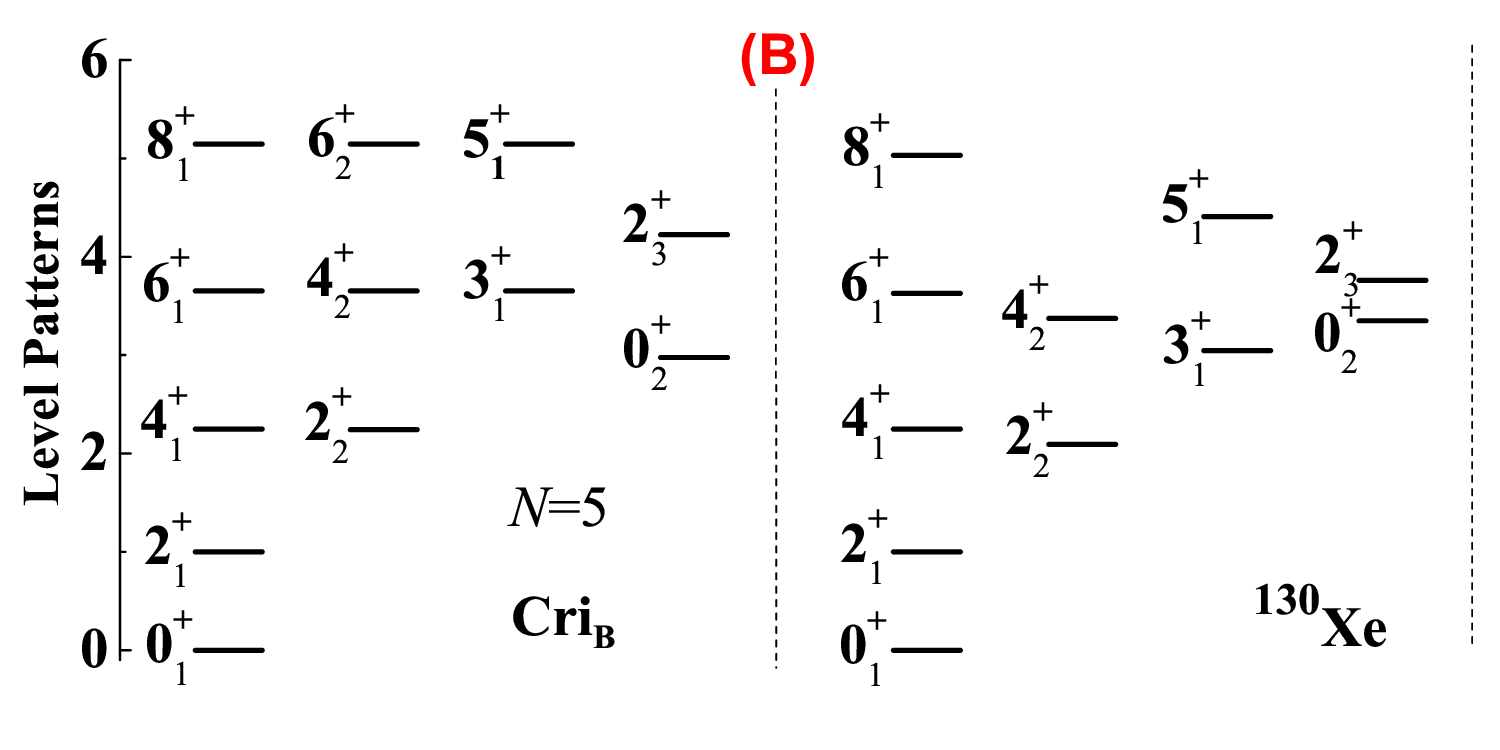}
\includegraphics[scale=0.35]{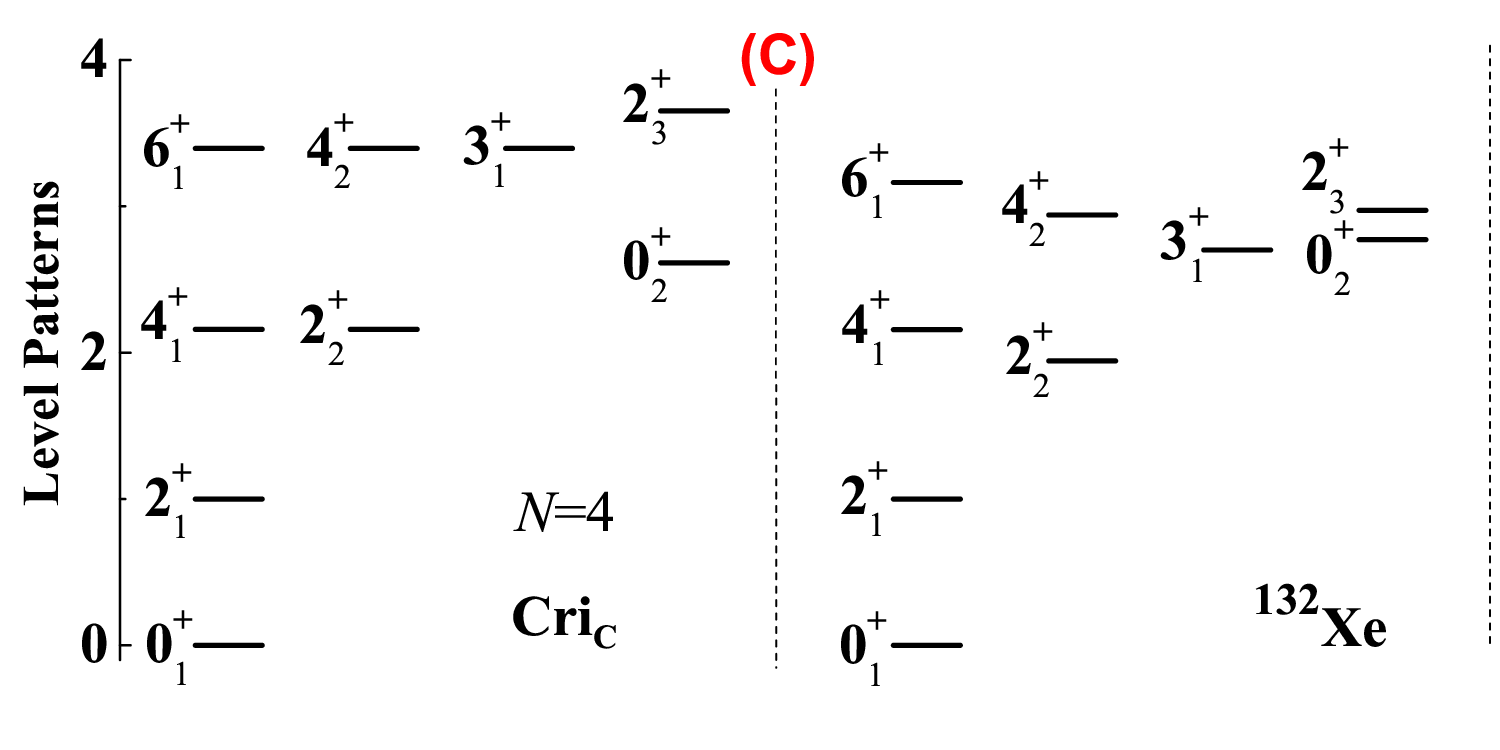}
\caption{(A) The low-lying level pattern (normalized to $E(2_1^+)=1.0$) for $^{128}$Xe~\cite{Elekes2015} and the corresponding critical point pattern ($N$=6), obtained from solving the Hamiltonian in (\ref{H}) with the parameter $\eta=0.88$ fixed to reproduce the experimental value of $R_{4/2}$. (B) The same as in (A), but for $^{130}$Xe~\cite{Singh2001} and the critical point pattern ($N$=5), solved using $\eta=0.82$. (C) The same as in (A), but for $^{132}$Xe~\cite{Khazov2005} and the critical point pattern ($N$=4), solved using $\eta=0.73$. \label{F5}}
\end{center}
\end{figure}

As discussed above, the critical point systems in the prolate-oblate SPT may exhibit spectroscopic quadrupole moment features resembling those of the Xe isotopes with $A\approx130$.
In the following, $^{128,130,132}$Xe~\cite{Elekes2015,Singh2001,Khazov2005} are selected as experimental counterparts for comparison with the critical structures predicted by the present SPT scheme.
Specifically, the available data for the low-lying patterns, typical $B(E2)$ values, and spectroscopic quadrupole moments for these three nuclei are presented in Fig.~\ref{F5}, Table~\ref{T1} and Table~\ref{T2}, respectively,
to compare with theoretical predictions derived from the Hamiltonian (\ref{H}). Since the focus is more on examining critical structure than on achieving a best fit to the data, the parameter $k$ in calculations is constantly fixed at $k_\mathrm{c}=3N/(2N+3)$, with boson number $N$ being set equal to the number of valence nucleon (hole) pairs, i.e., $N=6,~5$, and $4$ for $^{128}$Xe, $^{130}$Xe, and $^{132}$Xe, respectively. The solely remaining adjustable parameter $\eta$ in the Hamiltonian is simply determined by precisely reproducing corresponding empirical values of $R_{4/2}$, which are $2.33$, $2.25$ and $2.16$ for these isotopes. Notably, scaling factor $\varepsilon$ in the Hamiltonian does not influence dynamical structures and can be determined by reproducing excitation energy $E(2_1^+)$ in experiments when necessary. To test theoretical predictions on the $E2$ transitional properties, effective charge $e$ in the $E2$ operator is fully determined by reproducing experimental data for $B(E2;2_1^+\rightarrow0_1^+)$. The resulting values are approximately given by $e\simeq0.14$ eb for all three Xe isotopes, even though the experimental $B(E2;2_1^+\rightarrow0_1^+)$ values vary significantly among different nuclei.

\begin{table*}
\caption{The $B(E2)$ values (in W.u.) for $^{128,130,132}$Xe are presented in comparison with predications from the critical point (Cri.) systems, as well as results from the shell model (SM), the DF rotor model, the O(5)-CBS rotor, and the E(5) CPS. All values are normalized to the experimental data for $B(E2;2_1^+\rightarrow 0_1^+)$, except for the shell mode (SM) results, which are taken from \cite{Kisyov2022}, \cite{Kaneko2023} and \cite{Teruya2015} for $A=128$,$130$ and $132$, respectively. The experimental data are taken from \cite{Kisyov2022} for $^{128}$Xe, \cite{Morrison2020} for $^{130}$Xe, and \cite{Peters2019} for $^{132}$Xe, and the scripts "$a$" and "$b$" mean that the data are taken from \cite{Coquard2009} and \cite{Peters2016}, respectively. Parameters used in the different models are described in the text, and a dash "-" indicates that the result is either not available experimentally or absent in existing theoretical calculations.}\label{T1}
\begin{tabular}{ccccccc|cccccc|cccccc}\hline\hline
$L_i^\pi\rightarrow L_f^\pi)$&$^{128}$Xe&$\mathrm{\mathrm{Cri_A}}$&SM&DF&CBS&E(5)&$^{130}$Xe&$\mathrm{Cri_B}$&SM&DF&CBS&E(5)&$^{132}$Xe&$\mathrm{Cri_C}$&SM&DF&CBS&E(5) \\
\hline
$2_1^{+}\rightarrow0_1^{+}$&46.1(15)&46.1&44&46.1&46.1&46.1&32(3)&32&34&32&32&32&23.1(15)&23.1&27.7&23.1&23.1&23.1\\
$4_1^{+}\rightarrow2_1^{+}$&55.4(32)&63&66&65&72&77&47(4)&44&54&45&53&54&28.6(23)&31&40&32&38&39\\
$6_1^{+}\rightarrow4_1^{+}$&76(10)&69&80&82&92&102&60$_{-12}^{+14}$&45&68&56&71&71&140($_{-130}^{+150}$)&29&31&40&49&51\\
$2_2^{+}\rightarrow2_1^{+}$&44(4)&66&54&44&72&77&37(3)&44&48&37&53&54&41(4)&31&37&28&38&39\\
$2_2^{+}\rightarrow0_1^{+}$&0.58(9)&0.59&0.24&1.84&0&0&0.23(2)&0.16&0.2&0.74&0&0&0.079(11)&0.035&0.0002&0.079&0&0\\
$0_2^{+}\rightarrow2_1^{+}$&3.7(6)$^a$&16&-&-&0&40&18($_{-11}^{+17}$)$^b$&15&-&-&24&28&4.0($_{-29}^{+31}$)&16&-&-&18&20\\
$0_2^{+}\rightarrow2_2^{+}$&52.8(76)$^a$&108&-&-&92&0&120($_{-70}^{+110}$)$^b$&78&-&-&0&0&-&61&-&-&0&0\\
\hline\hline
\end{tabular}
\end{table*}

For comparative purposes, results for the relevant $B(E2)$ transitions and quadrupole moments extracted from other model calculations are also listed in Table~\ref{T1} and Table~\ref{T2}. These models include a recently established shell model method (called the PMMU model)~\cite{Kaneko2023,Kaneko2014}, the $\gamma$-rigid rotor model of DF~\cite{DF1958}, and two $\gamma$-soft rotor models, namely the well-known E(5) CPS~\cite{Iachello2000} and the O(5)-confined $\beta$-soft (O(5)-CBS) rotor~\cite{Bonatsos2006}.
Among these, the DF rotor and the two $\gamma$-soft rotors are considered here to represent two limiting cases of triaxiality: $\gamma$-rigid and $\gamma$-unstable. Specifically, the CBS model results are taken from Ref.~\cite{Bonatsos2006}, where the parameter values $r_\beta=0.21$ and $r_\beta=0.12$ were used in calculations for $^{128}$Xe and $^{130}$Xe, respectively. For $^{132}$Xe, the same value $r_\beta=0.12$ is assumed in extracting CBS model results. Note that all three Xe nuclei have ever been considered as potential candidates for comparison with solutions of the E(5) CPS~\cite{Peters2016,Clark2004,Coquard2009}, which were later extended to more general cases in the O(5)-CBS model~\cite{Bonatsos2006}. In contrast to the two $\gamma$-soft models, which may both yield vanishing spectroscopic quadrupole moments for all states, the $\gamma$-rigid rotor model can produce $\gamma$-dependent results for quadrupole moments. In the calculations~\cite{Li2016}, the $\gamma$ parameter in the DF rotor model is determined here by fitting experimental values of $E(2_2^+)/E(2_1^+)$, as this quantity is highly sensitive to $\gamma$ deformation in a $\gamma$-rigid rotor model. In this way, $\gamma=25.4^\circ, \gamma=26.8^\circ$, and $\gamma=28.9^\circ$ are obtained for $^{128}$Xe, $^{130}$Xe, and $^{132}$Xe, respectively. These results are in rough agreement with previous assumptions regarding $\gamma$ deformation of these nuclei~\cite{Kisyov2022}.

As seen from Fig.~\ref{F5}, the critical point level patterns obtained in the present model show generally good agreement with those of the low-lying states in $^{128,130,132}$Xe.
The theoretical results implicitly exhibit O(5)-like approximate degeneracies in level energies, such as $(4_1^+,~2_2^+)$ and $(6_1^+,~4_2^+,~3_1^+)$~\cite{IachelloBook}.
These degeneracies were also observed in a recent numerical analysis employing a similar Hamiltonian~\cite{Wang2022}, although a proper explanation has yet to be provided. In fact, the observed energy degeneracies can be partially explained in terms of the Hamiltonian confined at the critical points. As discussed above, the states belonging to the irreps $(\lambda,\mu)=(2N-2t,t)$ with $t=0,~2,~4,~\cdots$, generated by the SU(3) Hamiltonian (\ref{HSU3}) with $k=k_\mathrm{c}$, become the lowest-energy states and exhibit energy degeneracy due to level crossings at the critical point~\cite{Zhang2012}. Notably, these degenerate irreps $(\lambda,\mu)=(2N-2t,t),~t=0,~2,~4,~\cdots$, can form a special class of SU(3) configurations. Within this class, a subset of states with $K=t$ can be annihilated by the boson-pair operator $\hat{P}=\tilde{d}\cdot \tilde{d}-2s^2$, which acts as an SU(3) (2,0) tensor, as demonstrated in the analysis by Leviatan~\cite{Leviatan1996,Leviatan2011}. In other words, these degenerate SU(3) states $|\phi\rangle_{\mathrm{su3}}$ are zero-energy eigenstates of the Hamiltonian $\hat{H}_P=\hat{P}^\dag\hat{P}$, satisfying $\hat{H}_P|\phi\rangle_{\mathrm{su3}}=E|\phi\rangle_{\mathrm{su3}}$ with $E=0$. Since $\hat{H}_P$ is also an O(5) scalar~\cite{IachelloBook}, these zero-energy eigenstates arising from the degenerate SU(3) irreps may possess certain O(5) symmetry characteristics~\cite{Leviatan1985}, although these features remain obscured in the degenerate energy levels due to their zero-energy nature. When an O(5) scalar term such as $\hat{n}_d$ is added to the SU(3) Hamiltonian with $k=k_\mathrm{c}$ (the degenerate point), as implemented in the present scheme, the lowest states may acquire certain O(5) features~\cite{Leviatan2011}. This may, to some extent, account for the O(5)-like degeneracies observed in Fig.~\ref{F5}. While the inclusion of the $\hat{n}_d$ term is essential for generating such O(5)-like features, these approximate O(5)-like degeneracies may persist even when only a very small O(5) scalar component is introduced into the Hamiltonian (\ref{H}) at $k=k_\mathrm{c}$, which can induced only SU(3)-type degeneracies without any $L$-dependence. However, once the system deviates slightly away from the parameter $k=k_\mathrm{c}$, the O(5)-like characteristics vanish rapidly, accompanied by a sharp increase in the $R_{4/2}$ ratio, as illustrated in Fig.~\ref{F3}(e). This observation confirms that the O(5)-like degeneracies observed in Fig.~\ref{F5} are closely related to the corresponding degenerate SU(3) irreps. A more detailed investigation into the emergence of O(5)-like features from prolate-oblate phase transitions will be presented elsewhere.

It is noteworthy that such degeneracy features are approximately present in the low-lying spectra of these Xe nuclei, where the level pattern for $^{132}$Xe is truncated at $L^\pi=6^+$ due to the absence of data for higher spins. This observation is consistent with the original perspectives~\cite{Peters2016,Rainovski2010,Coquard2011} that the approximate O(5) symmetry may
be more or less preserved in Xe isotopes near $A=130$, which have been previously considered as candidates for various $\gamma$-soft models, such as the O(6) DS~\cite{Brentano1988}, the E(5) CPS~\cite{Clark2004}, or the O(5)-CBS model~\cite{Bonatsos2006}. Undoubtedly, introducing additional terms or relaxing the parameter constraints of the Hamiltonian may lift these degeneracies and enhance theoretical description; however, we retain the current scheme to preserve a simple geometric interpretation of the critical point system.

\begin{table*}
\caption{The available experimental data for the spectroscopic quadrupole moments (unit in eb) of low-spin states in $^{128,130,132}$Xe~\cite{Morrison2020,Kisyov2022} are presented and compared with theoretical predictions from various models, where the effective charge in the calculations has been determined by reproducing the experimental values of $B(E2;2_1^+\rightarrow0_1^+)$ transitions. Shell model results are taken from Ref.~\cite{Kisyov2022,Kaneko2023,Teruya2015}.
}\label{T2}
\begin{tabular}{ccccc|cccc|cccc}\hline\hline
$Q(L^+)$&$^{128}$Xe&$\mathrm{\mathrm{Cri_A}}$&SM&DF($25.4^\circ$)&$^{130}$Xe&$\mathrm{Cri_B}$&SM&DF($26.8^\circ$)&$^{132}$Xe&$\mathrm{Cri_C}$&SM&DF($28.9^\circ$)  \\
\hline\hline
$Q(2_{1}^{+})$&-0.44$_{-12}^{+9}$&-0.26&-0.37&-0.52&-0.38$_{-14}^{+17}$&-0.30&-0.26&-0.33&-&-0.31&-0.22&-0.11\\
$Q(4_{1}^{+})$&-1.04(10)&-0.61&-0.45&-0.31&-0.41(12)&-0.61&-0.32&-0.17&-&-0.63&-0.30&-0.05\\
$Q(2_{2}^{+})$&$\mathbf{0.008_{-0.08}^{+0.07}}$&$\mathbf{0.066}$&0.33&0.52&$\mathbf{0.1(1)}$&$\mathbf{0.11}$&0.25&0.33&-&$\mathbf{0.13}$&0.25&0.11\\
$Q(4_{2}^{+})$&-&-0.25&-&-0.98&-&-0.22&-&-0.81&-&-0.23&-&-0.44\\
\hline\hline
\end{tabular}
\end{table*}

Regarding $B(E2)$ transitions, the results in Table~\ref{T1} show that different models do not exhibit significant qualitative differences in describing most of the available $B(E2)$ data. These data are generally well accounted for by various theoretical calculations, including those obtained from the rigid DF rotor and E(5) CPS. However, for the weak $B(E2;0_2^+\rightarrow2_1^+)$ and strong $B(E2;0_2^+\rightarrow2_2^+)$ transitions observed experimentally~\cite{Peters2016,Coquard2009}, better agreement is found with the critical point calculations and those from the O(5)-CBS model with a relatively large $r_\beta$ value~\cite{Bonatsos2006}. These results deviate from those predicted by the E(5) model or the O(5)-CBS model with small $r_\beta$ value (see the case for $^{130}$Xe). Note that the DF model cannot generate a $0_2^+$ state due to its assumption of freezing the $\beta$ degree of freedom~\cite{DF1958}. Furthermore, finite-$N$ effects can be observed in the critical point results for small $N$, when compared to other collective models. For instance, the critical point calculation with $N=4$ predicates $B_{6/2}=B(E2;6_1^+\rightarrow4_1^+)/B(E2;4_1^+\rightarrow2_1^+)<1.0$, as shown in Table~\ref{F1}. This feature aligns with the shell model result~\cite{Teruya2015} and contrasts with predictions from other collective models, all of which yield $B_{6/2}>1.0$. Nevertheless, experimental verification of this behavior remains challenging due to large uncertainties in corresponding data for $^{132}$Xe.

Since the effective charges used in the calculations have been determined by reproducing the experimental data for $B(E2;2_1^+\rightarrow0_1^+)$, a more stringent test of the theoretical predictions from different models can be obtained from the spectroscopic quadrupole moments, which
provide the most decisive information for nuclear shapes~\cite{Kaneko2023}.
As shown in Table~\ref{T2}, two key features can be identified from the experimental data for spectroscopic quadrupole moments: $Q(4_1^+)<Q(2_1^+)<0$ and $0<Q(2_2^+)\ll |Q(2_1^+)|$.
The first feature is well reproduced by results from the critical point calculations and shell model results, whereas it deviates to varying degrees from predictions based on the rigid rotor calculations. The latter typically yields $0>Q(4_1^+)>Q(2_1^+)$ and $0>Q(2_1^+)=-Q(2_2^+)$ for $\gamma<30^\circ$. The second feature also aligns well with critical point predictions, particularly for $^{128}$Xe, where the nearly vanishing $Q(2_2^+)$ has previously posed a challenge to theoretical explanation~\cite{Kisyov2022}. Similarly, the critical point results for $^{130}$Xe appear to be in better agreement with experimental data than those from other models, especially concerning the prediction of an unusually small $Q(2_2^+)$ value.
Furthermore, current calculations predict similar features for spectroscopic quadrupole moments in $^{132}$Xe, which are not yet available experimentally, thus providing a basis for future measurements. In addition, a comparable amplitude with $Q(4_2^+)\sim Q(2_1^+)$ has been predicted across all considered Xe isotopes, in contrast to the rigid rotor model predictions, which typically yield larger values with $|Q(4_2^+)|\geq2|Q(2_1^+)|$, as shown in the Table. This highlights a new characteristic that can be used to distinguish between $\gamma$-rigid and $\gamma$-soft models. Overall, the present calculations based on prolate-oblate SPT scheme offer further insights into these unusual quadrupole moments observed in Xe nuclei near $A=130$.

\begin{table}
\caption{The calculated values (in degrees) of $\overline{\gamma}_\mathrm{su3}$ and $\Delta\overline{\gamma}_\mathrm{su3}$ for various nuclear states are presented using the same parameters as those in Fig.~\ref{F5}. Here, $\overline{\gamma}_c$ and $\Delta\overline{\gamma}_c$ represent the results taken from ref.~\cite{Bally2022}, where the average $\gamma$ deformation and its fluctuation for the ground states of $^{128,130}$Xe were evaluated based on the collective wave function obtained from the multi-reference energy density functional calculations.}

\label{T3}
\begin{tabular}{c|cccc|cccc|cc}\hline\hline
$L^{\pi}$&&$^{128}$Xe&&&&$^{130}$Xe&&&$^{132}$Xe\\
&$\overline{\gamma}_{\mathrm{su3}}$&$\Delta\overline{\gamma}_{\mathrm{su3}}$&$\overline{\gamma}_c$&$\Delta\overline{\gamma}_c$&$\overline{\gamma}_{\mathrm{su3}}$&$\Delta\overline{\gamma}_{\mathrm{su3}}$&$\overline{\gamma}_c$&$\Delta\overline{\gamma}_c$
&$\overline{\gamma}_{\mathrm{su3}}$&$\Delta\overline{\gamma}_{\mathrm{su3}}$\\
\hline
$0_1^{+}$&$\mathbf{25^{\circ}}$&$\mathbf{14^{\circ}}$&$\mathbf{21^{\circ}}$&$\mathbf{13^{\circ}}$&$\mathbf{24^{\circ}}$&$\mathbf{13^{\circ}}$&$\mathbf{23^{\circ}}$&$\mathbf{16^{\circ}}$&$\mathbf{25^{\circ}}$&$\mathbf{15^{\circ}}$\\
$2_1^{+}$&$23^{\circ}$&$14^{\circ}$&&&$22^{\circ}$&$12^{\circ}$&&&$21^{\circ}$&$14^{\circ}$\\
$4_1^{+}$&$19^{\circ}$&$13^{\circ}$&&&$18^{\circ}$&$12^{\circ}$&&&$16^{\circ}$&$12^{\circ}$\\
$2_2^{+}$&$24^{\circ}$&$14^{\circ}$&&&$24^{\circ}$&$13^{\circ}$&&&$26^{\circ}$&$16^{\circ}$\\
$0_2^{+}$&$21^{\circ}$&$19^{\circ}$&&&$18^{\circ}$&$16^{\circ}$&&&$18^{\circ}$&$20^{\circ}$\\
\hline\hline
\end{tabular}
\end{table}

As discussed above, the exotic spectroscopic quadrupole moments observed in experiments can be well reproduced by the current theoretical calculations. This suggests that these Xe nuclei are situated near the critical point of the prolate-oblate SPT, and therefore, their low-lying states are expected to possess considerable $\gamma$-softness. This picture of $\gamma$ deformation is qualitatively consistent with previous mean-field studies~\cite{Abolghasem2020}, as well as with recent Hartree-Fock-Bogolyubov calculations on the Xe isotopes given in \cite{Kaneko2023}. To provide a clearer understanding of the $\gamma$ deformation extracted from the current calculations, the average $\gamma$ deformation and its fluctuation for low-spin states are listed in Table~\ref{T3}.
As observed in Table~\ref{T3}, relatively large $\gamma$ deformations, along with significant fluctuations, are obtained for all low-spin states in the current model. In particular, the calculated values of $\overline{\gamma}_\mathrm{su3}$ and $\Delta\overline{\gamma}_\mathrm{su3}$ for $0_1^+$ are very close to the results derived from the multi-reference energy density functional method, as given in \cite{Bally2022}, confirming that these Xe nuclei are indeed $\gamma$-soft.
Furthermore, the average $\gamma$ deformation and its fluctuation in yrast states are found to decrease slightly with increasing spin, but remain substantially large. Notably, the results in Table~\ref{T3} show that the $\gamma$ deformations for $2_1^+$ and $2_2^+$ are nearly identical across all three Xe isotopes despite significant differences in their corresponding spectroscopic quadrupole moments observed in Table~\ref{T2}. This implies that the critical point systems corresponding to $^{128,130,132}$Xe are effectively "stabilized" in a $\gamma$-soft situation where an unusually small $Q(2_2^+)$ naturally arises.

It should be noted that a sign change between $Q(2_1^+)$ and $Q(2_2^+)$, as shown in Table~\ref{T2}, does not necessarily imply that the system undergoes a change in deformation across different states, especially when $2_2^+$ corresponds to the $\gamma$-band head. This point can be clarified by examining the DF rotor calculations provided in Table~\ref{T2}, which clearly demonstrate that the rotor model predicts $Q(2_2^+)=-Q(2_1^+)$ for any $\gamma$-deformation. However, since critical behaviors arise only during the prolate to oblate SPT in the present model, these exotic quadrupole moments may reflect an intrinsic instability of the prolate shape against oblate deformation. Therefore, the current results support the recent prolate-oblate analysis of these Xe nuclei based on shell model calculations~\cite{Kaneko2023}. That analysis revealed that the microscopic $QQ$ interaction acting on $\Delta j=2$ oribis $(1h_{11/2},2f_{7/2})$ within the quasi-SU(3) coupling scheme can drive the nuclear shape from oblate to prolate at the critical neutron number $N_\mathrm{n}=76$. In addition, although three Xe isotopes near $N_\mathrm{n}=76$ are all described by critical systems in the present calculations, they exhibit different contributions from the U(5) mode, corresponding to different $\eta$ values adopted in Hamiltonian. This picture aligns with the traditional IBM perspective on the U(5)-O(6) transitions during structural evolution along the Xe isotopes chain~\cite{Casten1985,Bonatsos2008}. In fact, within the Casten triangle, the parameter points lying on the U(5)-O(6) SPT path correspond precisely to the critical points in the SU(3)-$\overline{\mathrm{SU(3)}}$ description of the prolate-oblate SPT~\cite{Jolie2001,Jolie2002}, which has been investigated through shell model calculations~\cite{Kaneko2023} based on the quasi-SU(3) coupling scheme~\cite{Zuker1995,Zuker2015} as mentioned above.

\begin{center}
\vskip.2cm\textbf{V. Summary}
\end{center}\vskip.2cm

In summary, an algebraic scheme of the prolate to oblate SPT is proposed within the IBM~\cite{IachelloBook} to elucidate the exotic quadrupole structures observed in Xe isotopes near $A=130$.
The resulting model, with parameters constrained at the critical point, offers a reasonable explanation of the experimental data on low-lying states.
Particularly, the results demonstrate that extremely small values of $Q(2_2^+)$, which posed a challenge~\cite{Kisyov2022} in understanding the nuclear structures of these Xe nuclei, naturally emerge from the critical point systems in the present prolate-oblate SPT scheme. It is further revealed that considerable $\gamma$-softness may occur in such systems, as evidenced by large
$\gamma$ deformation and fluctuations across different nuclear states. Therefore, the present calculations provide a simple yet effective perspective on the quadrupole deformation of these Xe isotopes, which exhibit unusual spectroscopic quadrupole moments~\cite{Morrison2020,Kisyov2022}. Additionally, the present scheme offers an alternative framework for describing prolate-oblate SPTs in nuclei, beyond the conventional SU(3)-O(6)-$\overline{\mathrm{SU(3)}}$ description~\cite{Jolie2001}. This may open new avenues for the microscopic understanding of nuclear shapes induced by nucleons involved in the quasi-SU(3) coupling scheme in shell model calculations~\cite{Kaneko2023}. It would also be valuable to examine the present SPT scheme in other nuclei with soft triaxial deformations~\cite{Jolie2003}. The related work is in progress.

\bigskip

\begin{acknowledgments}
YZ gratefully acknowledges Yang Sun for numerous constructive discussions on the related topics as well as for bringing our attention to the unusual quadrupole moments observed in Xe isotopes.
This work was supported by the National Natural Science Foundation of China (12375113).
\end{acknowledgments}

\end{document}